# Rate-equation calculations of the current flow through two-site molecular device and DNA-based junction

Kamil Walczak

Institute of Physics, Adam Mickiewicz University
Umultowska 85, 61-614 Poznań, Poland

Here we present the calculations of incoherent current flowing through the two-site molecular device as well as the DNA-based junction within the rate-equation approach. Few interesting phenomena are discussed in detail. Structural asymmetry of two-site molecule results in rectification effect, which can be neutralized by asymmetric voltage drop at the molecule-metal contacts due to coupling asymmetry. The results received for poly(dG)-poly(dC) DNA molecule reveal the coupling- and temperature-independent saturation effect of the current at high voltages, where for short chains we establish the inverse square distance dependence. Besides, we document the shift of the conductance peak in the direction to higher voltages due to the temperature decrease.

Key words: incoherent transport, DNA-based device, DNA electronics, rate equation, molecular diode

PACS numbers: 72.10.Bg, 73.40.Ei, 85.65.+h, 87.15.Aa

## 1. Introductory remarks

The fundamental role of DNA molecule, or deoxyribonucleic acid, is to carry the genetic code of organisms. Such biomolecule is made of a sequence of four bases: thymine (T), cytosine (C), adenine (A), and guanine (G), attached to a phosphate-sugar backbone. Any particular sequence forms a single strand of DNA. Its double-helix structure, as discovered by Watson and Crick [1], is composed of two strands coupled together by hydrogen bonds between A and T as well as G and C bases. Anyway, transport and self-assembly properties of biomolecules have recently attracted considerable attention because of their possible applications in future electronic devices [2-6]. DNA itself could be suitable for charge conduction, because some of the orbitals belonging to the bases overlap quite well with each other along the axis of the molecular chain.

The actual magnitude of DNA conductivity and its physical mechanism is still under debate, since experiments indicate that the discussed biomolecule can behave as: an Anderson insulator [7-11], a wide-band-gap semiconductor [12-15], an Ohmic conductor [16-25] or even a proximity-induced superconductor [26]. Among all the conditions that can have significant influence on the experimental results we can enumerate: DNA base sequence, molecular length, orientation in space, stretching, nature of the molecule-electrode contact, ambient surrounding (humidity), temperature of the system, adsorption surface, structural form of the molecule (bundle or single chain), and preparation conditions. In particular, long DNA oligomers are expected to be true insulators, while short biomolecules are rather semiconductors with a relatively large highest occupied molecular orbital – lowest unoccupied molecular orbital (HOMO-LUMO) gap ~few eV.

In the experiment conducted by Porath *et al.* [12], nonlinear current-voltage (I-V) characteristics of a single 10-nm-long poly(dG)-poly(dC) DNA molecule with a homogenous sequence (30 GC pairs) connected to Pt electrodes were measured directly, using scanning tunneling microscope (STM apparatus). These results suggest that the transport characteristics reflect the electronic structure of the molecular bands of the DNA [27,28]. Besides, it is believed that the charge transfer is due to incoherent hole-type (positive ion) transport between energetically appropriate HOMO levels of guanines (nucleobases with lower oxidation potential). Here decoherence is understood in the sense that the phase of charge carriers is lost as a result of their temporarily localization on guanine molecule.

A quantitative account of the mentioned results has been provided by Li and Yan [29] by means of a simple homogenous one-band tight-binding model for HOMO-mediated charge transport and dephasing treated within the Büttiker's approach, where the phase-breaking processes are modeled by coupling each GC pair to fictitious electronic reservoir. Later on, Zwolak and Di Ventra using the same method have predicted that spin-valve behavior can be observed in experiments with ferromagnetic electrodes [30]. The main purpose of this work is to perform the calculations within the rate-equation approach [31-35] in order to study the influence of few important factors on incoherent transport through two-site molecular device as well as DNA-based junction, namely: (i) the strength of the molecule-electrode coupling, (ii) asymmetry of the contact, (iii) the length of DNA molecule, and (iv) the temperature of the system under investigation. It should be also noted that the mentioned method was used earlier to calculate numerically the $I-V$ dependences of DNA-based devices [36-39].

## 2. Transport through two-site molecule

Molecular diodes have always played a key role in the development of molecular electronics as a proposal of the simplest components in future electronic circuits [40]. Rectifying behavior of molecular devices, where the magnitude of the junction current strongly depends on bias polarity, can be explained as a consequence of two dominant factors: structural asymmetry of molecular complex [41,42], and asymmetric potential spatial profile along the junction due to asymmetry in molecule-metal contacts [43]. Here we show that the combined effect of these two factors can lead to very small or no rectification at all, since they can cancel out each other.

Let us consider molecule composed of two different subunits (atoms, benzene rings with different chemical substituents) separated by insulating bridge ($\sigma$-bonded) and weakly connected to two identical electrodes via potential barriers. In our model, molecular subunits are treated as quantum dots with discrete energy levels, while connecting bridge is modeled as a potential barrier for moving electrons. The existence of tunnel barriers in molecular junctions can be questionable, however such barriers at a molecule-metal interfaces have been reported in the literature [31,44]. Furthermore, we restrict our calculations to the highest occupied molecular orbitals of particular subunits (HOMO levels), assuming that the other energy levels are beyond the transport window and hybridization effects can be neglected.

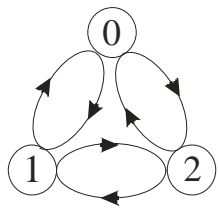

Figure 1: Schematic diagram of the two-site model device, where all the possible transitions between reservoir (0) and two molecular subunits (1,2) are depicted.

Incoherent electron transport is described as sequential tunneling events of a charge carrier between the electrodes (reservoir 0) and molecular subunits (1 and 2), as shown in Fig.1. To determine the probabilities of finding an electron on each distinguished site, we can write down the following rate equations:

$$\dot{P}_0(t) = -[\gamma_{0\to1} + \gamma_{0\to2}]P_0(t) + \gamma_{1\to0}P_1(t) + \gamma_{2\to0}P_2(t), \tag{1}$$

$$\dot{P}_1(t) = -[\gamma_{1\to0} + \gamma_{1\to2}]P_1(t) + \gamma_{0\to1}P_0(t) + \gamma_{2\to1}P_2(t), \tag{2}$$

$$\dot{P}_2(t) = -[\gamma_{2\to0} + \gamma_{2\to1}]P_2(t) + \gamma_{0\to2}P_0(t) + \gamma_{1\to2}P_1(t). \tag{3}$$

Description in terms of rate equations is adequate when relaxation processes on the particular subunits (1 and 2) are much faster than the time scale for tunneling between them. It should be mentioned that some of the experimental data may be interpreted as incoherent tunneling [31,45]. Limiting ourselves to steady state conditions, where $\dot{P}_0(t) = \dot{P}_1(t) = \dot{P}_2(t)$, we have:

$$[\gamma_{0\to1} + \gamma_{0\to2}]P_0 = \gamma_{1\to0}P_1 + \gamma_{2\to0}P_2, \tag{4}$$

$$[\gamma_{1\to0} + \gamma_{1\to2}]P_1 = \gamma_{0\to1}P_0 + \gamma_{2\to1}P_2, \tag{5}$$

$$[\gamma_{2\to0} + \gamma_{2\to1}]P_2 = \gamma_{0\to2}P_0 + \gamma_{1\to2}P_1. \tag{6}$$

Out of the above three equations only two of them are linearly independent and therefore in order to determine all these stationary probabilities, we have to take advantage of the so-called normalization condition:

$$P_0 + P_1 + P_2 = 1. \tag{7}$$

It is easy to solve Eqs.(4)-(7) and express the occupation probabilities with the help of the proper transition rates between particular subunits:

$$P_0 = \frac{1}{M_I}[\gamma_{1\to2}\gamma_{2\to0} + \gamma_{1\to0}(\gamma_{2\to0} + \gamma_{2\to1})], \tag{8}$$

$$P_1 = \frac{1}{M_I}[\gamma_{0\to2}\gamma_{2\to1} + \gamma_{0\to1}(\gamma_{2\to0} + \gamma_{2\to1})], \tag{9}$$

$$P_2 = \frac{1}{M_I}[\gamma_{0\to1}\gamma_{1\to2} + \gamma_{0\to2}(\gamma_{1\to0} + \gamma_{1\to2})], \tag{10}$$

where denominator is given by:

$$M_I = (\gamma_{1\to0} + \gamma_{1\to2})(\gamma_{0\to2} + \gamma_{2\to0}) + \gamma_{2\to1}(\gamma_{0\to2} + \gamma_{1\to0}) + \gamma_{0\to1}(\gamma_{1\to2} + \gamma_{2\to0} + \gamma_{2\to1}). \tag{11}$$

To obtain all the probabilities, we have to know the transition rates that can be determined by assuming a specific model for the coupling between all the model sites. Using Fermi golden-rule arguments we can write down the following expressions [35]:

$$\gamma_{0\to i} = \frac{\Gamma_i}{\hbar}\int_{-\infty}^{+\infty}d\varepsilon D(\varepsilon - \varepsilon_i)f(\varepsilon - \mu_i), \tag{12}$$

$$\gamma_{i \to 0} = \frac{\Gamma_i}{\hbar} \int_{-\infty}^{+\infty} d\varepsilon D(\varepsilon - \varepsilon_i)[1 - f(\varepsilon - \mu_i)], \tag{13}$$

where: $i = 1, 2$. Here $\Gamma_1$ and $\Gamma_2$ are parameters describing the strength of the coupling with the source and drain electrodes, $f(\varepsilon - \mu_i) = [\exp[\beta(\varepsilon - \mu_i)] + 1]^{-1}$ denotes Fermi distribution function defined by the following electrochemical potentials [43]:

$$\mu_1 = \varepsilon_F + (1 - \eta)eV, \tag{14}$$

$$\mu_2 = \varepsilon_F - \eta eV. \tag{15}$$

Here: $\beta = (k_B T)^{-1}$, $k_B$ is the Boltzmann constant, $T$ is the temperature of the system under investigation, $\varepsilon_F$ is the Fermi level, while the voltage division factor: $\eta = \Gamma_2 / (\Gamma_1 + \Gamma_2)$. The broadening of molecular energy levels is taken into account through densities of states of a Lorentzian shape (by assumption):

$$D(\varepsilon - \varepsilon_i) = \frac{\Gamma_i / 2\pi}{(\varepsilon - \varepsilon_i)^2 + \Gamma_i^2 / 4}. \tag{16}$$

The charge transfer rate between the localized donor and acceptor sites is mediated by off-resonance superexchange coupling [46]:

$$\gamma_{1 \leftrightarrow 2} = \frac{1}{\tau} \exp[-\alpha R], \tag{17}$$

where: $\tau$ is the tunneling time, $\alpha$ is a structure-dependent attenuation factor, $R$ is the distance between two molecular species (the length of a bridge). The time $\tau$ is approximately equal to experimentally-determined time involved into the bridge vibrations and generally can be temperature-dependent. For saturated hydrocarbon chains $\alpha \approx 0.9$ Å$^{-1}$ [46], while the calculated distance decay exponent for the hole transfer between GC pairs $\alpha = 0.95 - 1.5$ Å$^{-1}$ [47]. Besides, assuming $R = 6$ Å and molecular vibration period for that bridge of order of $\tau \sim 10^{-12}$ s [48], the transfer rate is $\gamma_{1 \leftrightarrow 2} \approx 0.005$ eV. Electrical current flowing through the analyzed system can be written in the form:

$$I = 2e[\gamma_{0 \to 1} P_0 - \gamma_{1 \to 0} P_1] = \frac{2e}{M_I}[\gamma_{0 \to 1} \gamma_{1 \to 2} \gamma_{2 \to 0} - \gamma_{1 \to 0} \gamma_{2 \to 1} \gamma_{0 \to 2}]. \tag{18}$$

Factor of 2 in Eq.(18) accounts for two spin orientations.

Figure 2 presents the calculated $I - V$ dependences for a two-site molecular complex (with different site energy levels) connected to metallic electrodes. Here we show the results obtained for the case of symmetric and asymmetric coupling at room temperature, where all the model parameters are chosen to show that the combined effect of structural and coupling asymmetries can provide almost symmetric $I - V$ function. From the enclosed picture we can see that diode-like behavior is significantly reduced by asymmetric voltage drop at the molecule-metal contacts due to coupling asymmetry.

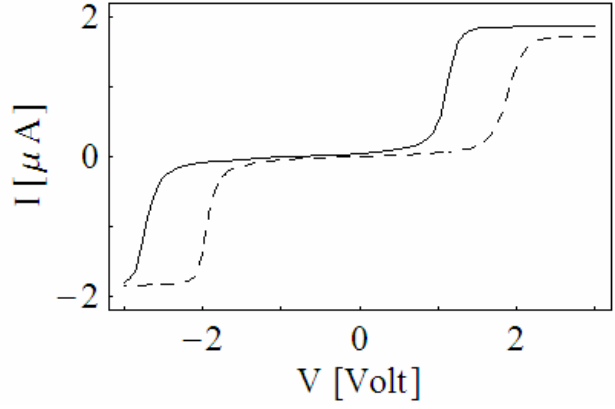

Figure 2: The current-voltage characteristics for the two-site molecular device in the case of symmetric ($\Gamma_1 = \Gamma_2 = 0.05$: solid line) and asymmetric coupling ($\Gamma_1 = 0.07$, $\Gamma_2 = 0.03$: dashed line) with the electrodes, where: $\varepsilon_1 = -0.4$, $\varepsilon_2 = 0.4$, $\varepsilon_F = 1.0$, $\gamma_{1\leftrightarrow 2} = 0.005$, $\beta^{-1} = 0.025$. All the model parameters are given in eV.

## 3. Transport through DNA molecule

To proceed, let us assume that the molecule, connected at both ends to metallic electrodes (reservoir 0), is represented by a set of $N$ guanine-cytosine pairs ($1,\ldots,N$), as shown in Fig.3. As mentioned in introduction, here we take into consideration one HOMO level of the energy $\varepsilon_H$ on each GC pair. The hole-type transport is treated as incoherent process, and therefore the time evolution of the occupation probabilities on particular model sites can be expressed by a set of the following $N+1$ rate equations:

$$\dot{P}_0(t) = -[\gamma_{0\to 1} + \gamma_{0\to N}]P_0(t) + \gamma_{1\to 0}P_1(t) + \gamma_{N\to 0}P_N(t), \qquad (19)$$

$$\dot{P}_1(t) = -[\gamma_{1\to 0} + \gamma_{1\to 2}]P_1(t) + \gamma_{0\to 1}P_0(t) + \gamma_{2\to 1}P_2(t), \qquad (20)$$

$$\ldots$$

$$\dot{P}_i(t) = -[\gamma_{i\to i-1} + \gamma_{i\to i+1}]P_i(t) + \gamma_{i-1\to i}P_{i-1}(t) + \gamma_{i+1\to i}P_{i+1}(t), \qquad (21)$$

$$\ldots$$

$$\dot{P}_N(t) = -[\gamma_{N\to N-1} + \gamma_{N\to 0}]P_N(t) + \gamma_{N-1\to N}P_{N-1}(t) + \gamma_{0\to N}P_0(t). \qquad (22)$$

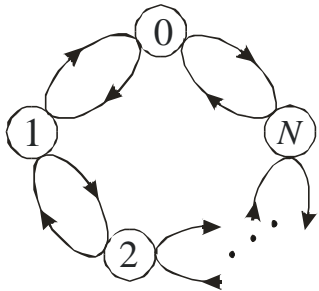

Figure 3: Schematic diagram of the DNA model junction, where all the incoherent transitions between reservoir ($0$) and the GC subunits of DNA molecule ($1,\ldots,N$) are depicted.

Since we are interested in the steady-state conditions, $\dot{P}_i(t) = 0$ for $i = 0,...,N$:

$$[\gamma_{0\to1} + \gamma_{0\to N}]P_0 = \gamma_{1\to0}P_1 + \gamma_{N\to0}P_N, \qquad (23)$$

$$[\gamma_{1\to0} + \gamma_{1\to2}]P_1 = \gamma_{0\to1}P_0 + \gamma_{2\to1}P_2, \qquad (24)$$

$$\ldots$$

$$[\gamma_{i\to i-1} + \gamma_{i\to i+1}]P_i = \gamma_{i-1\to i}P_{i-1} + \gamma_{i+1\to i}P_{i+1}, \qquad (25)$$

$$\ldots$$

$$[\gamma_{N\to N-1} + \gamma_{N\to 0}]P_N = \gamma_{N-1\to N}P_{N-1} + \gamma_{0\to N}P_0, \qquad (26)$$

Of course, the normalization condition still should be fulfilled:

$$\sum_{i=0}^{N} P_i = 1. \qquad (27)$$

Again, the transfer rates between end GC subunits and the electrodes can be evaluated through the use of the Fermi golden-rule:

$$\gamma_{0\to1} = \frac{\Gamma_1}{\hbar} \int_{-\infty}^{+\infty} d\varepsilon D_1(\varepsilon - \varepsilon_H) f(\varepsilon - \mu_1), \qquad (28)$$

$$\gamma_{1\to0} = \frac{\Gamma_1}{\hbar} \int_{-\infty}^{+\infty} d\varepsilon D_1(\varepsilon - \varepsilon_H)[1 - f(\varepsilon - \mu_1)], \qquad (29)$$

$$\gamma_{0\to N} = \frac{\Gamma_2}{\hbar} \int_{-\infty}^{+\infty} d\varepsilon D_2(\varepsilon - \varepsilon_H) f(\varepsilon - \mu_2), \qquad (30)$$

$$\gamma_{N\to0} = \frac{\Gamma_2}{\hbar} \int_{-\infty}^{+\infty} d\varepsilon D_2(\varepsilon - \varepsilon_H)[1 - f(\varepsilon - \mu_2)], \qquad (31)$$

where densities of states are given by:

$$D_x(\varepsilon - \varepsilon_H) = \frac{\Gamma_x/2\pi}{(\varepsilon - \varepsilon_H)^2 + \Gamma_x^2/4}, \qquad (32)$$

for $x = 1,2$. Besides, for simplicity, let us treat the transition rates between two neighboring GC pairs as energy- and bias- independent parameters that can be expressed with the help of constant value:

$$\gamma_{i\leftrightarrow i+1} = \gamma_M = const., \qquad (33)$$

for $i = 1,...,N-1$. To simplify our notation we introduce the following ratios:

$$\xi_L \equiv \frac{\gamma_M}{\gamma_{0\to1}}, \qquad (34)$$

$$\xi_R \equiv \frac{\gamma_M}{\gamma_{N \to 0}}, \tag{35}$$

$$\chi_L \equiv \frac{\gamma_{1 \to 0}}{\gamma_{0 \to 1}}, \tag{36}$$

$$\chi_R \equiv \frac{\gamma_{0 \to N}}{\gamma_{N \to 0}}. \tag{37}$$

Simple transformation brings our Eqs.(23)-(26) to the form:

$$[\xi_R + \chi_R \xi_L] P_0 = \chi_L \xi_R P_1 + \xi_L P_N, \tag{38}$$

$$[\chi_L + \xi_L] P_1 = P_0 + \xi_L P_2, \tag{39}$$

...

$$2 P_i = P_{i-1} + P_{i+1}, \tag{40}$$

...

$$[1 + \xi_R] P_N = \xi_R P_{N-1} + \chi_R P_0, \tag{41}$$

Now we have to deal with homogenous set of equations, where among all the $N+1$ equations only $N$ are linearly independent, so we have to complete Eqs.(38)-(41) by normalization condition given by Eq.(27). Eq.(40) represents the molecular part and due to a two-step recursion method can be expressed in terms of $P_1$ and $P_N$:

$$P_i = \frac{N-i}{N-1} P_1 + \frac{i-1}{N-1} P_N. \tag{42}$$

Using the above Eq.(42) in order to determine $P_2$ and $P_{N-1}$ and inserting such relations to the Eqs.(39) and (41) we can write down a set of three equations (together with normalization):

$$[\chi_L + \xi_L] P_1 = P_0 + \xi_L \left[ \frac{N-2}{N-1} P_1 + \frac{1}{N-1} P_N \right], \tag{43}$$

$$[1 + \xi_R] P_N = \xi_R \left[ \frac{1}{N-1} P_1 + \frac{N-2}{N-1} P_N \right] + \chi_R P_0, \tag{44}$$

$$P_0 + \frac{N}{2} [P_1 + P_N] = 1. \tag{45}$$

Solving the Eqs.(43)-(45) we obtain the probabilities useful in the current calculations:

$$P_0 = \frac{2}{M_{II}} [\xi_L + \chi_L (\xi_R + N - 1)], \tag{46}$$

$$P_1 = \frac{2}{M_{II}} [\xi_L \chi_R + \xi_R + N - 1], \tag{47}$$

$$P_N = \frac{2}{M_{II}}[\xi_R + \chi_R(\xi_L + \chi_L(N-1))], \qquad (48)$$

where denominator is given by:

$$M_{II} = 2[\xi_L + \chi_L(\xi_R - 1)] + N[\chi_R(2\xi_L - \chi_L) + 2(\xi_L + \xi_R) - 1] + N^2[1 + \chi_L\chi_R]. \qquad (49)$$

In the standard rate-equations approach, the electrical current flowing through the junction can be calculated as:

$$I = 2e[\gamma_{0\to 1}P_0 - \gamma_{1\to 0}P_1]. \qquad (50)$$

The differential conductance is then given by the derivative of the current with respect to voltage.

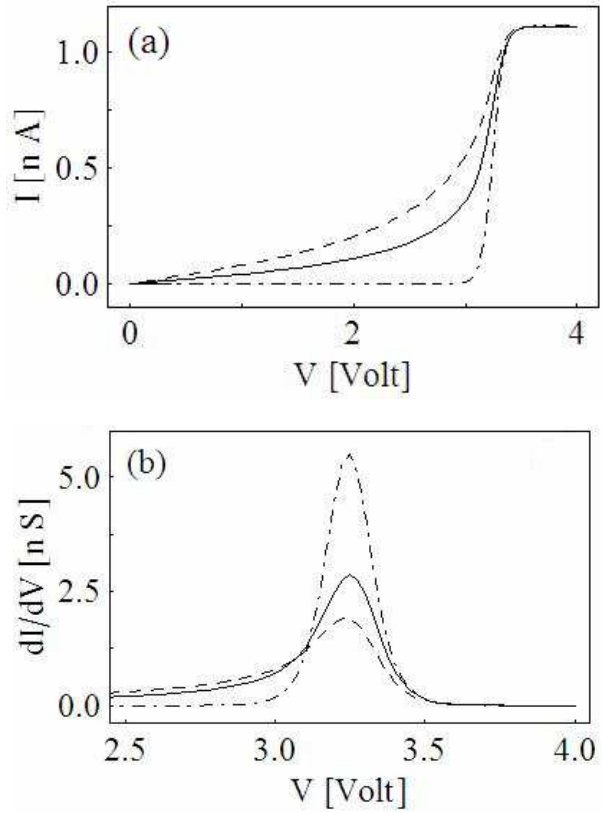

Figure 4: The current-voltage (a) and conductance-voltage characteristics (b) for DNA molecule composed of $N = 30$ GC pairs connected to metallic electrodes calculated for the case, where the coupling-induced broadening is neglected ($\Gamma_1 = \Gamma_2 = 0.01$: dashed-dotted line) and for two cases, where the broadening effect is included ($\Gamma_1 = \Gamma_2 = 0.01$: solid line; $\Gamma_1 = \Gamma_2 = 0.015$: dashed line). Here: $\varepsilon_H = 0$, $\varepsilon_F = 1.73$, $\gamma_M = 0.001$, $\beta^{-1} = 0.025$. All the model parameters are given in eV.

The calculated $I - V$ characteristics for a double-stranded 30-base pair poly(dG)-poly(dC) DNA molecule connected to two metallic electrodes are demonstrated in Fig.4a. Here we show the results obtained for the case of symmetric coupling at room temperature, where all the model parameters are chosen to quantitatively reproduce the experimental data [12]. Ignoring the coupling-induced broadening (where $D_x(\varepsilon - \varepsilon_H) \approx \delta(\varepsilon - \varepsilon_H)$), the distinct step-like behavior of the current curve is predicted. Anyway, in our model, we can observe the saturation effect at higher voltages right after the jump in the $I-V$ dependence, where we have $dI/dV = 0$. As expected, the mentioned broadening effect results in smoothing of the current curve. But, interestingly, an increase of the coupling parameter has no influence on the maximal value of the current flowing through the junction. Therefore, increasing the strength of the molecule-electrode coupling, the height of the peak in the differential conductance spectrum is reduced, while the peak broadening increases, as shown in Fig.4b.

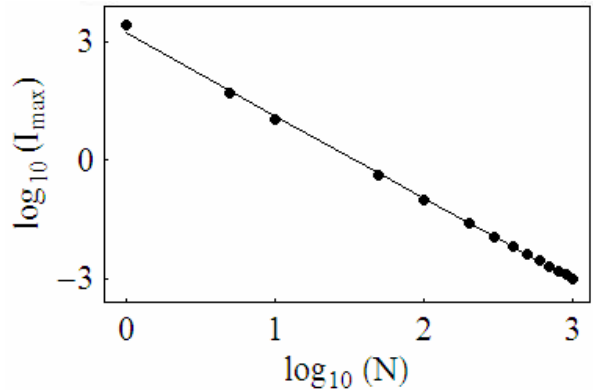

Figure 5: Maximal value of the current flowing through DNA chain ($I_{max} = I(V=4)$) as a function of the number of GC subunits ($N$), where: $\varepsilon_H = 0$, $\varepsilon_F = 1.73$, $\Gamma_1 = \Gamma_2 = 0.01$, $\gamma_M = 0.001$, $\beta^{-1} = 0.025$. All the model parameters are given in eV.

It should be noted that increasing the length of DNA by attaching another GC pairs to the main chain we can observe very large reduction of the maximal bias-independent current and insignificant shift of the current step in the direction to lower voltages. Figure 5 demonstrates strong dependence of the saturation current from the length of the DNA molecule. Our calculations indicate that the magnitude of the current can change in a wide range of values – from $\sim mA$ for very short DNA fragments ($\sim 1nm$) to $\sim pA$ for longer DNA chains ($\sim 0.5\mu m$). For the model parameters chosen in this section, the following proportion is fulfilled:

$$\log_{10}(I_{max}) \cong 3.2 - 2\log_{10}(N).  \qquad (51)$$

The magnitude of the maximal current given in nA can be estimated from the relation:

$$I_{max} \cong 1585 \times N^{-2}. \qquad (52)$$

The maximal current for shorter DNA chains reveals stronger algebraic length dependence than in the case of longer biomolecules [39]. Generally, we write down the following relation: $I_{max} \propto N^{-\phi}$, where for the case of shorter DNA chains ($N < 1000$) we have the inverse square distance regime and $\phi = 2$, while for the case of very long DNA molecules we have the inverse distance regime and $\phi = 1$ (as appropriate for diffusional charge transport regime). Moreover, we have studied the influence of the thermal effects on the changes in transport characteristics. It was found that the maximal value of the current is temperature independent, while decreasing the junction temperature we observe the shift of the conductance peak in the direction to higher voltages, as documented in Fig.6.

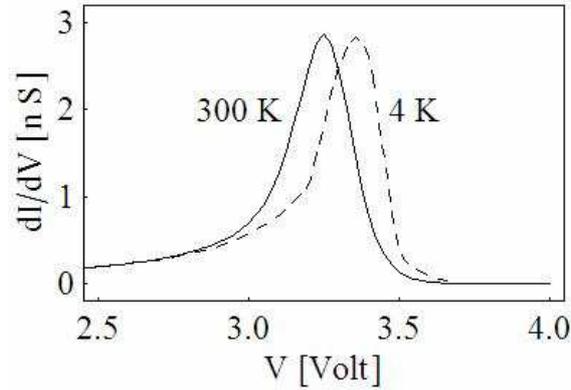

Figure 6: Differential conductance as a function of bias voltage for DNA molecule composed of $N = 30$ GC pairs connected to metallic electrodes calculated for two different temperatures: $T = 300$ K ($\beta^{-1} = 0.025$: solid line) and $T = 4$ K ($\beta^{-1} = 0.00035$: dashed line). The other model parameters are the same as in Fig.5.

Finally, it should be mentioned about further perspectives in modeling the transport phenomena in DNA-based devices. In the incoherent transport regime, the charge transferred between particular GC pairs is temporarily localized on guanine. If the time spending on the G part is comparable or longer than the tunneling time between GC subunits, Coulomb interactions between charge carriers can modify the current flowing through DNA molecule [34]. Such eventuality is not taken into account in our model. Besides, the charge transfer is very sensitive to the motion of base pairs, since structural changes can lead to charge localization. Transport can also have polaron character due to the coupling of the electronic and vibrational-mode degrees of freedom [49]. Moreover, transport mechanism related to metal-DNA-metal junctions can be associated with soliton formation due to the formation of domain walls in dimerized bonds of DNA [50].

The unique assembly features of DNA molecule, its recognition ability, optical properties, flexibility, and adaptability suggest that DNA may become one of the most important species in molecular electronics and biotechnology. In particular, DNA has potential for assembling networks with variety of geometries, such as: loops, arrays, cubes, knots, and nanolattices [51,52].

**Acknowledgement**

The author is very grateful to M. Sidowski and K. Herbeć for illuminating discussions.